\documentclass[fleqn,10pt]{wlscirep}
\title{Emergent stochastic oscillations and signal detection in tree networks of excitable elements}

\def\mean#1{\langle#1\rangle}
\newcommand{\Dunits}{($\mu$A/cm$^2$)$^2$\text{ms}}
\newcommand{\constCurrent}{I}
\newcommand{\IAH}{I_{\text{AH}}}
\newcommand{\ISN}{I_{\text{SN}}}
\newcommand{\Ith}{I_{\text{th}}}
\newcommand{\Ieff}{I_{\text{eff}}}
\newcommand{\Deff}{D_{\text{eff}}}

\author[1]{Justus Kromer}
\author[2]{Ali Khaledi-Nasab}
\author[3,4]{Lutz Schimansky-Geier}
\author[2,5,*]{Alexander B. Neiman}

\affil[1]{Center for Advancing Electronics Dresden, TU Dresden, Mommsenstrasse 15, 01069 Dresden, Germany}
\affil[2]{Department of Physics and Astronomy, Ohio University, Athens, Ohio 45701, USA}
\affil[3]{Department of Physics, Humboldt-Universit\"at zu Berlin, Newtonstrasse 15, 12489 Berlin, Germany}
\affil[4]{Bernstein Center for Computational Neuroscience, Berlin, Germany}
\affil[5]{Neuroscience Program, Ohio University, Athens, Ohio 45701, USA}
\affil[*]{neimana@ohio.edu}

\begin{abstract}
We study the stochastic dynamics of strongly-coupled excitable elements on  a tree network. 
The peripheral nodes receive independent random inputs which may induce large spiking events 
propagating through the branches of the tree and leading to global coherent oscillations in the network. 
This scenario may be  relevant to action potential generation in certain sensory neurons, 
which possess myelinated distal dendritic tree-like arbors with excitable nodes of Ranvier at 
peripheral and branching nodes and exhibit noisy periodic sequences of action potentials.
We focus on the spiking statistics of the central node, which fires in response to a noisy 
input at peripheral nodes. We show that, in the strong coupling regime, relevant to myelinated dendritic trees, 
the spike train statistics can be predicted from an isolated excitable element with rescaled parameters according to 
the network topology. Furthermore, we show that by varying the network topology the spike train statistics 
of the central node can be tuned  to have a certain firing rate and variability, or 
to allow for an optimal discrimination of inputs applied at the peripheral nodes.   
\end{abstract}

\begin{document}
\flushbottom
\maketitle
\thispagestyle{empty}

\section*{Introduction}
Coupled noisy excitable systems serve as relevant models for a wide range of natural 
phenomena, including pattern 
formation in chemical reactions \cite{kiss2003chemical,mikhailov2012engineering} and 
in social networks \cite{farkas2002social,newman2002spread,perc2006coherence,borgatti2009network}, 
dynamics of gene regulatory networks \cite{chen2015emergent} and of single and 
networked neurons \cite{Copelli2005691,furtado2006response,maxim2015}.   
Networks of noisy excitable elements exhibit a rich variety of spatio-temporal 
dynamics, depending on the strength and topology of coupling and the noise intensity  
\cite{jung1995spatiotemporal,lin2000,sagues2007spatiotemporal}. For example, the coherence of emergent 
network oscillations can be controlled by modifying the noise intensity, the coupling 
strength, or by changing the network size or topology 
\cite{toral2003,perc2005spatial,perc2007stochastic,pregosak2010,kal2010,son2013}. 
The dynamic range and sensitivity of complex networks of excitable elements to 
external stimuli can by optimized for critical topologies 
\cite{gollo2013single,kinouchi2006optimal,larremore2011predicting}.
\begin{figure}[h!]
 \centering
 \includegraphics[width=0.5\linewidth]{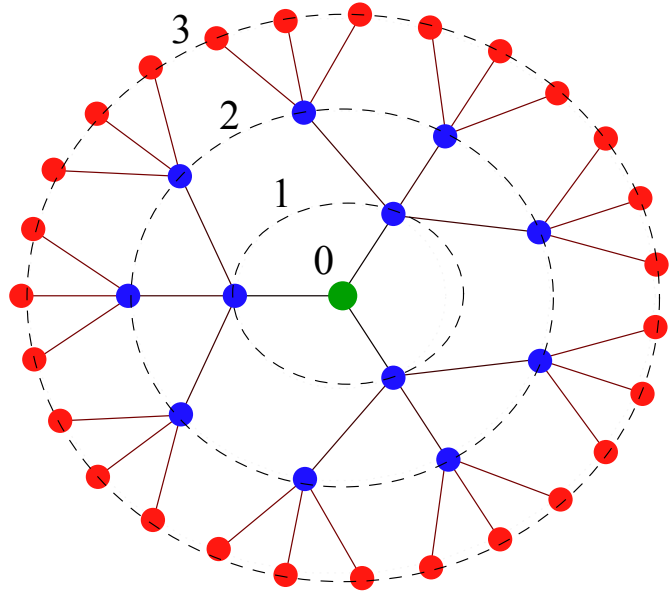}
\caption{Tree network with branching order, $d=3$, and $G=3$ generations. Peripheral nodes are 
marked red and receive external excitations. Dashed circles indicate corresponding shells of the tree's generations, $g=1$, 2 and 3; generation $g=0$ refers to the central (primary branching) node (green). For a discrete cable model of myelinated dendrites,  active elements are nodes of Ranvier, which are connected by passive resistors.}
\label{f1}
\end{figure}

In the present paper, we focus on the
dynamics of regular tree networks of strongly coupled excitable elements which receive random and independent excitations to their peripheral nodes, as sketched in Fig.~\ref{f1}.
Our study is motivated by the morphology of certain peripheral sensory neurons, which
possess branched myelinated dendritic terminals at their receptive 
fields, with multiple nodes of Ranvier. Their extended terminal branching resembles the dendrite structure of 
neurons in the central nervous system (CNS) \cite{london2005dendritic,st2016}. Myelinated segments form a tree-like structure with nodes of Ranvier at each branching point.  
Myelination terminates at peripheral nodes of Ranvier, called heminodes, which receive sensory signals.  
Thus, such sensory neurons may possess multiple spike initiation zones at  
heminodes which encode a local sensory signal into a stream of action potentials (APs)
which are merged into a single output spike train transmitted to the CNS \cite{eagles1974afferent}.
Examples for such neurons are the afferent innervation of muscle spindles 
\cite{quick1980anatomical,banks1982form,banks1997pacemaker,banks2009comparative}, pain
receptors \cite{besson1999neurobiology}, cutaneous mechanoreceptors 
\cite{lesniak2014computation,walsh2015mammalian}, and lung receptors \cite{lee2014sensory}. 
Interestingly, sensory neurons with myelinated dendrites may exhibit spontaneous activity 
characterized by 
coherent periodic spiking, despite that their peripheral heminodes presumably receive 
uncorrelated noisy excitations \cite{banks1997pacemaker}. 
Figure \ref{f1} then can be viewed as a model for a branched myelinated 
dendritic terminal, where peripheral nodes receive uncorrelated stochastic inputs 
and are linked by myelinated segments. 
Due to the high density of Na$^+$ ion channels at the nodes of Ranvier, 
APs may be excited independently at different peripheral nodes. 
 High electrical conductivity of myelinated 
segments, which link the individual nodes, result in a strong coupling between the nodes. 
Therefore, their stochastic dynamics synchronizes. This may result 
in noisy periodic spiking of the primary branching (central) node, as we have shown for star networks of excitable elements \cite{justus2016}.

Here we use a biophysical model for nodes of Ranvier connected 
by myelinated links on regular trees and show numerically and analytically that 
the collective response of the network can be deduced from the stochastic dynamics of a single effective node 
with parameters scaled according to the network size and topology.
Thus, our study allows for the prediction of the stochastic network dynamics from the tree topology.
We then discuss how the tree topology affects the firing statistics of the central node and the discriminability of input signals.

\section*{Model and Methods}
\subsection*{Discrete Cable Model}
In the present paper, we study the stochastic 
dynamics of excitable elements linked  on  a regular tree (see 
Fig.~\ref{f1}). Branching starts at the primary (central) node (number 0 in Fig.~\ref{f1}) 
and continues through several generations.  Only the peripheral nodes receive external inputs.
Referring to a model of branched myelinated dendrites, these peripheral nodes are called 
heminodes and receive inputs from thin unmyelinated processes (neurites).  
APs are initiated at the heminodes and then propagate on the tree towards the primary 
branching node and eventually to the CNS.

Here we consider regular trees whose topology is characterized by two parameters: the branching, $d$, and the number of generations, $G$. 
Given these two parameters, the total number of nodes, $N$, and the number of peripheral nodes, $N_p$, are given by 
\begin{equation}
N = \frac{d^{G+1}-1}{d-1} \ \quad \text{and} \ \quad N_p=d^G,
\label{model.eq1}
\end{equation}
respectively. The dynamics of the membrane potential is approximated by a discrete cable model 
\cite{ermentrout2010foundations} in which nodes of Ranvier are connected by passive 
resistive links according to the network topology. All active nodes and passive links 
are assumed to be identical, except that peripheral nodes receive external inputs. 
The membrane potential $V_k(t)$ of the $k$th node obeys the dynamics
\begin{eqnarray}
\label{model.eq2}
&&C \dot{V}_k = - I_\text{ion} [V_k,\mathbf{u}_k] + \kappa \sum_{j=0}^{N-1} A_{k,j}(V_j - V_k) + I_{{\rm ext}_k}, \\
&&\dot{\mathbf{u}}_k = \mathbf{\alpha}(V_k)(1-\mathbf{u}_k)-\mathbf{\beta}(V_k)(1-\mathbf{u}_k), \nonumber 
\end{eqnarray}
where the index $k=0,1,2,...,N-1$ marks the respective node. In particular, $k=0$ refers to the central node. 
In Eq. (\ref{model.eq2}) the term $I_{ion}[V_k, {\bf u}_k]$ stands for nodal ionic 
currents and ${\bf u}_k(t)$ is a vector whose components are the gating variables of the nodal ion 
channels and $C=2~\mu$F/cm$^2$ is the nodal capacitance per area. 
In the following we use two particular models for the nodes of Ranvier: 
a Hodgkin-Huxley-type (HH) model with Na$^+$ and leak currents \cite{justus2016} 
and the Frankenhaeuser-Huxley (FH) model which includes 
additional K$^+$ and persistent Na$^+$ currents. The HH nodal model includes
two gating variables, $m$ and $h$, for Na$^+$ channels, i.e. $I_\text{ion}[V,\mathbf{u}]=I_\text{ion}[V,m,h]$. The FH model includes two additional gating variables, $n$ for K$^+$, and $p$ for persistent Na$^+$ channels:
$I_\text{ion}[V,\mathbf{u}]=I_\text{ion}[V,m,h,n,p]$.  The detailed equations and 
parameters of the nodal models are provided in the Supplementary Material.

The coupling term in Eq. (\ref{model.eq2}), $\kappa \sum_{j=0}^{N-1} A_{k,j}(V_j - V_k)$,
contains the adjacency matrix $\mathbf{A}$ of the tree graph and the coupling 
strength, $\kappa$, in units of Siemens per area. Its value can be calculated from the sizes 
of the node and myelinated links, and the axoplasmic resistivity:  
\begin{equation}
\kappa = \frac{a}{4 l L \rho},
\label{model.eq3} 
\end{equation}
where $a$ is the diameter of the node (and of links), $l$ is the nodal length, $L$ is 
the length of connecting links and $\rho$ is the axoplasmic resistivity. 
For example, for $\rho=100$~$\Omega$cm,  the nodal diameter and 
length $a=10$~$\mu$m, $l=1$~$\mu$m, and the length of myelinated 
segment $L=200$~$\mu$m, the coupling strength is $\kappa=1250$~mS/cm$^2$. 
This provides a biophysically-plausible range for $\kappa$, which we use as a control parameter in the following.

The external current $I_{ext}$ is applied only to the peripheral nodes and consists of a 
constant part $\constCurrent$ and noisy part, i.e. 
\begin{equation}
\label{model.eq4}
I_{{\rm ext}_k} = \delta_{k,p} [\constCurrent +\sqrt{2D} \, \xi_p(t)],
\end{equation}
where $p$ denotes indicies of peripheral nodes;  
$\delta_{k,p}$ is the Kronecker delta;
$D$ scales the intensity of the Gaussian white noise $\xi_p(t)$, which is uncorrelated for 
different peripheral nodes,
$\mean{\xi_{i}(t)\xi_{j}(t+\tau)}=\delta_{i,j} \,\delta(\tau)$. 
Thus, peripheral nodes receive random uncorrelated inputs. 

Equations (\ref{model.eq2}) were integrated numerically using explicit Euler–Maruyama methods with timestep of 0.1$\mu$s.

\subsection*{Variability of Generated Sequences of Action Potentials}
Our primary interest is the statistics of a spike train generated by the central node. A spike is identified as a full-size AP with a magnitude of at least 60~mV.  We extracted a sequence of spike times, $t_j$, at the central node from 60 -- 120~s long simulation runs. The corresponding sequence of interspike  intervals (ISIs) $\Delta t_j =t_{j+1}-t_j$, 
is characterized by the mean firing rate, $r$ and the coefficient of variation, $C_V$ as,
\begin{equation}
r=\frac{1}{\langle \Delta t_j \rangle}, \quad
C_V=r \sqrt{\langle (\Delta t_j -\langle \Delta t_j \rangle )^2\rangle},
\label{ee2}
\end{equation}
where the average is taken over all ISIs in the spike train of the central node.

\subsection*{Signal Detection}
To characterize the signal detection capacity of a tree network, we 
considered a small constant stimulus, $\Delta \constCurrent$, applied to the peripheral 
nodes in addition to the stimulus $\constCurrent$, and calculated a 
normalized distance between resulting spike count distributions of the central node with and without this addition. Such a measure of distance is given by the discriminability, $d'$, defined as \cite{stemmler1996},
\begin{equation}
d'=2\frac{|\mu_T(\constCurrent)-\mu_T(\constCurrent+\Delta \constCurrent)|}{\sigma_T(\constCurrent)+\sigma_T(\constCurrent+\Delta \constCurrent)},
\label{discrim.eq1}
\end{equation}  
where $\mu_T$ and $\sigma_T$ are the mean and standard deviation of the spike count in a time interval $T$, respectively. The discriminability quantifies how well the network 
responses to two different stimuli,
$\constCurrent$ and $\constCurrent+\Delta \constCurrent$, can be distinguished by observing corresponding spike count statistics at the central node.

The discriminability is related to the Fisher information, which provides the theoretical limit of how accurately a stimulus $\constCurrent$ can be estimated by observing a spike train \cite{cover2012}. 
For the spike count statistics, a lower bound of the Fisher information can be written as \cite{stemmler1996},
\begin{equation}
J_\text{LB}(\constCurrent)=\frac{1}{\sigma^2_T(\constCurrent)}\,\left(\frac{d \mu_T}{d\constCurrent}\right)^2,
\label{fisher.eq}
\end{equation}  
and is related to the discriminability, $d'$ by \cite{stemmler1996},
\begin{equation}
d'\approx \Delta \constCurrent \sqrt{J_\text{LB}(\constCurrent)}.
\label{dscrim.eq2}
\end{equation}
Larger values of the Fisher information refer to more accurate estimation of the stimulus from the spike train and so to better discrimination between two stimuli $\constCurrent$ and $\constCurrent+\Delta \constCurrent$.

The discriminability Eq. (\ref{discrim.eq1}) was calculated by collecting spike counts of the central node for $5000$ independent time intervals of lengths $T=200$~ms, and calculating the mean and standard deviation for 
two values of the stimulus, $\constCurrent$ and $\constCurrent+\Delta \constCurrent$, applied to the peripheral nodes of a tree network \cite{stemmler1996}. We also calculated the lower bound of the Fisher information Eq. (\ref{fisher.eq}) for the single uncoupled node as a function of the input current (stimulus) $\constCurrent$ and the noise intensity, $D$, using a similar numerical procedure.

\section*{Results}

\subsection*{Emergence of Periodic Firing in Deterministic Tree Networks}
At first, we consider the case of a deterministic input, $D=0$.
In the absence of the external input, $I_{\rm ext}=0$, an isolated node is in the excitable regime. 
A sufficiently high constant current, $\IAH$ results in a subcritical 
Andronov-Hopf bifurcation of the equilibrium state rendering an isolated node to 
fire a periodic sequence of APs. 
The corresponding limit cycle disappears in a saddle-node bifurcation for a lower 
external current, $\ISN$. For the HH nodal model the saddle-node bifurcation occurs at $\ISN \approx 28.15 \mu$A/cm$^2$ 
and the subcritical Andronov-Hopf bifurcation at $\IAH\approx 29.06 \mu$A/cm$^2$, so in a narrow 
range $\ISN<\constCurrent<\IAH$ an isolated node is bistable, possessing a stable equilibrium and a stable 
limit cycle.  When the nodes are coupled on a tree network and external currents are applied 
to the peripheral nodes, the dynamics of the network may become quite complex. For example, 
in case of weak coupling, peripheral nodes fire APs, which fail 
to propagate to the central node, so that nodes in the inner generations of the network exhibit 
small-amplitude spikes. For a stronger coupling, nodes in the inner generations may fire APs, but with skipping relative to APs in the periphery, demonstrating various $m:n$ synchronization patterns. 
However, for strong coupling and sufficiently high external 
currents the network shows fully synchronized periodic firing.

\begin{figure}[h]
\centering
\includegraphics[width=0.7\linewidth]{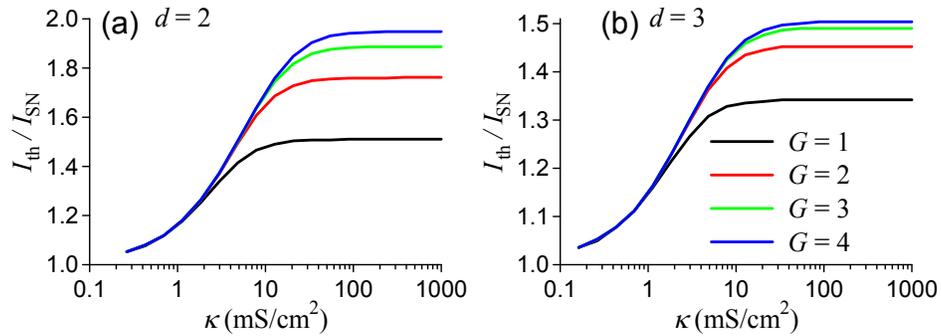}
\caption{Threshold constant current  $\Ith$  vs the coupling 
strength, $\kappa$, which gives rise to repetitive APs at the central node of the tree 
network with branching $d$ and the number of generation, $G$. The threshold 
current is normalized by
the bifurcation value $\ISN=28.15~\mu$A/cm$^2$ at which the single isolated HH node 
starts to generate a periodic sequence of APs.
Colors (see legends) refer to the indicated number of generations, $G$. Panels 
correspond to trees with branching $d=2$ (a) and $d=3$ (b), respectively.} 
\label{det1.fig}
\end{figure}

A comprehensive analysis of the deterministic dynamics is beyond the scope of this study. 
Instead, since our primary interest is in the emergence of periodic sequences of full-size 
APs at the central node, we address the following question:  Given the tree topology, 
$G$ and $d$, and the coupling strength, $\kappa$, what is a threshold value $\constCurrent^{\text{th}}$ 
of a constant current applied to peripherals, $\constCurrent$, which makes the central node 
to generate repetitive firing of full-size APs?
To this end, we perform simulations of tree networks with given $\kappa$, $G$ and $d$. Initially 
membrane potentials of the individual nodes are randomly distributed around the stable equilibrium 
of an isolated node for $\constCurrent=0$. Then we apply a current $\constCurrent > 0$ and 
determined the minimal value, $\constCurrent^{\text{th}}$ of $\constCurrent$
at which the central node generated APs repetitively at steady state.
Results are shown in Fig.~\ref{det1.fig}.
At the central node periodic firing of APs occurs for values of $\constCurrent$ and $\kappa$ above 
the corresponding curves in Fig.~\ref{det1.fig}. Below these curves, the 
network is excitable in the sense that no repetitive firing of APs is observed at the central node.
In the following, we refer to these two regimes as oscillatory (repetitive firing of full-size 
APs by the central node) and excitable  (no repetitive firing of APs by the central node).
The threshold value of the external current, $\constCurrent^{\text{th}}$, increases for 
weak and moderate values of the coupling strength. Consequently the network needs stronger 
external input to the peripheral nodes to sustain periodic 
firing of the central node. 

\begin{figure}[h]
\centering
\includegraphics[width=0.4\linewidth]{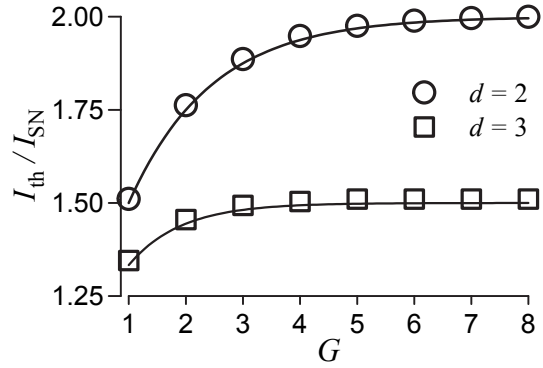}
\caption{Threshold constant current $\Ith$ vs. the number of generations for strong coupling 
$\kappa=100$~mS/cm$^2$ and indicated values of branching, $d$.
The threshold current is normalized by the bifurcation value $\ISN=28.15~\mu$A/cm$^2$ 
at which a single isolated HH node starts to fire a periodic sequence of APs.
Symbols indicate results from simulations and solid lines show the scaling relation
 Eq. (\ref{eq:scalingThresholdCurrent}).} 
\label{det2.fig}
\end{figure}

Figure ~\ref{det1.fig} shows two distinct coupling regimes. For 
weak coupling, $\kappa < 2$~mS/cm$^2$, the threshold current $\constCurrent^{\text{th}}$ is 
independent of the network size, i.e. the number of generations, $G$, and branching, $d$. 
In contrast, for strong coupling, $\kappa >60$~mS/cm$^2$, the threshold current saturates,
and its value increases with increasing number of generations. This is 
illustrated further in Fig.~\ref{det2.fig} showing the threshold current vs
the number of generations for strong coupling. Note that the 
strong coupling regime spans the range of realistic coupling strengths for models of branched myelinated 
dendrites. As can be seen in Fig.~\ref{det2.fig}, the threshold current follows a characteristic dependence 
saturating for trees with a large number of generations, $G$, and decreases 
with the increase of branching, $d$. Apparently, this dependence follows the scaling relation: 
\begin{eqnarray}
\label{eq:scalingThresholdCurrent}
\Ith/I_\text{b}=R^{-1}(G,d),
\end{eqnarray} 
where $I_\text{b}$ is a bifurcation value of the constant current in the isolated single node and the scaling factor $R(G,d)$ is the ratio of the number of 
peripheral nodes to the total number of nodes,
\begin{equation}
R(G,d) = \frac{N_p}{N} = \frac{d^G(d-1)}{d^{G+1}-1},
\label{r.eq}
\end{equation}
whith $N_p = d^G$ and $N=\sum_{k=0}^G d^k$, for regular trees. 
This scaling relation, Eq. (\ref{eq:scalingThresholdCurrent}), holds 
for strongly-coupled trees and is derived below. 
Deterministic trees with the FH nodal model show similar dynamics with the same 
scaling as in Fig.~\ref{det2.fig}.

\subsection*{Stochastic Dynamics}
\begin{figure}[h]
\centering
\includegraphics[width=0.75\linewidth]{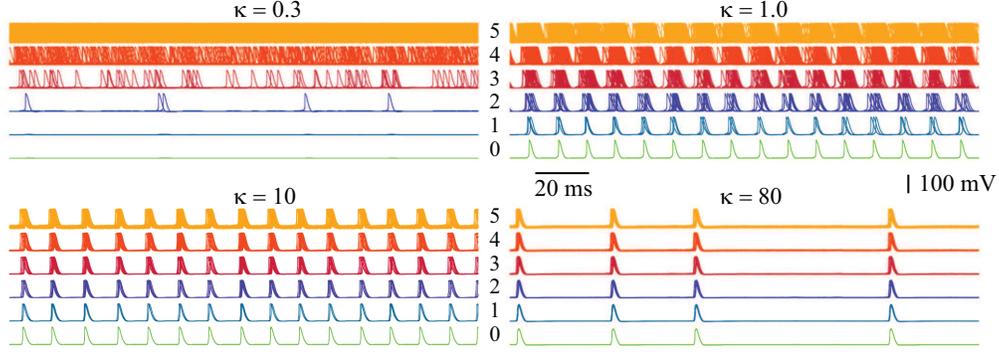}
\caption{Voltage traces of the 
HH nodes for a tree with $G=5$ generations and $d=3$ branching in the excitable regime. 
Peripheral nodes were excited by the external currents (\ref{model.eq4}) with parameters
$\constCurrent = 20~\mu A$/cm$^2$ and $D=500$~\Dunits. Each panel shows $200$~ms long 
superimposed voltage traces of nodes within a generation $g$, $g=1,..,5$, for the indicated 
coupling strength, $\kappa$ (in mS/cm$^2$). Horizontal axis is time. Numbers next to 
voltage traces indicate generations within the tree, $g=0$ corresponds to the central node and 
$g=5$ corresponds to the peripheral generation, respectively.
}
\label{stoch1.fig}
\end{figure}

The addition of uncorrelated noise to the peripheral nodes allows for the 
generation of APs in the excitable regime. Fig.~\ref{stoch1.fig} shows an example of the 
stochastic dynamics for a tree with
$G=5$ generations and $d=3$ branching. In the excitable regime ($\constCurrent = 20~\mu$A/cm$^2$) noise 
of sufficient intensity induces APs in peripheral nodes. For weak coupling 
($\kappa=0.3$~mS/cm$^2$) noise-induced APs in adjacent generations 
%the peripheral generation and generation next to it 
are not synchronized (superimposed spikes for peripheral nodes fill densely corresponding 
generation panels) and do not propagate beyond the 2-nd generation, which shows only sparse APs. 
Increasing the coupling strength leads to progressive synchronization of nodes in adjacent generations 
and finally results in the generation of APs in the central node. For strong coupling the whole network fires 
almost in synchrony. We note, however, that even for strong coupling outer 
generations show some spike jitter.
We also note that strong coupling leads to slower and more random firing of APs. 

\begin{figure}[h]
\centering
\includegraphics[width=0.7\linewidth]{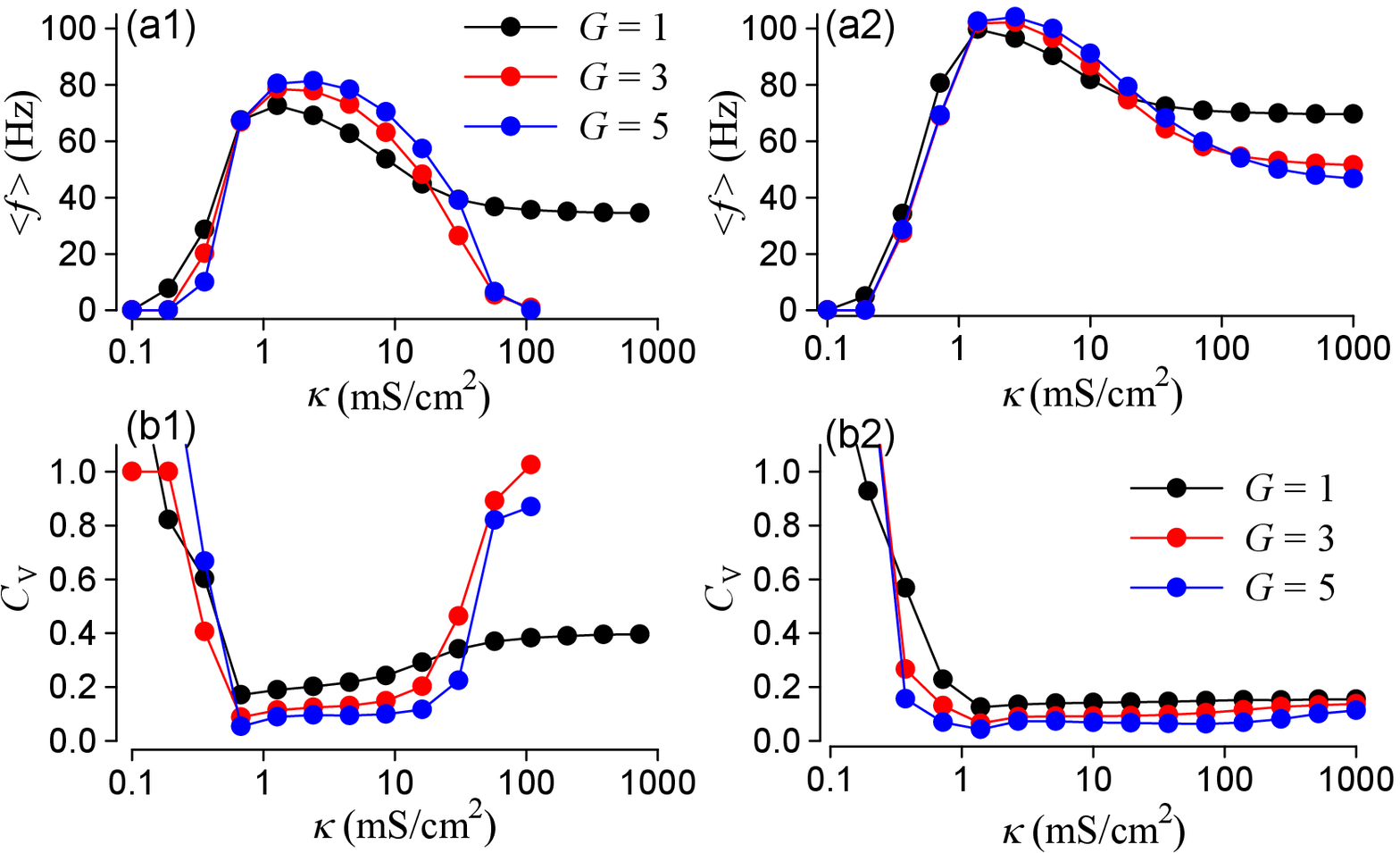}
\caption{Mean firing rate (a1, a2) and coefficient of variation of interspike 
intervals (b1, b2) of the central node versus coupling strength for tree 
networks with $d=2$ branching for the indicated numbers of generations, $G$. 
Left panels (a1, b1) correspond to the excitable regime with the constant current $\constCurrent=20~\mu$A/cm$^2$; right panels (a2, b2) refer to the oscillatory regime with $\constCurrent=60~\mu$A/cm$^2$. 
The noise intensity is $D=500$~\Dunits. 
}
\label{stoch2.fig}
\end{figure}

As observed for star networks\cite{justus2016}, the dynamics of the central node in a tree 
network depends non-monotonously on the 
coupling strength.  As shown in Fig.~\ref{stoch2.fig}, there exist 
optimal, rather small values of the coupling strength for excitable and oscillatory trees 
at which fastest (maximum firing rate) and most coherent (minimal coefficient of variation, $C_V$) firing is observed, respectively.
For extremely weak coupling APs, which are fired by different peripheral nodes, 
are not synchronized and fail to propagate to the central node (Fig.~\ref{stoch1.fig}, upper left panel). 
Increasing the coupling strength leads to stronger interaction between 
the branch nodes and results in synchronous  firing of all nodes.   

However, the size of a tree, i.e. the number of generations, is critical for the firing 
statistics of the central node. Furthermore, excitable and oscillatory trees demonstrate 
qualitatively different behaviour in the biologically-relevant strong coupling regime.
In excitable trees, firing of APs becomes slower and more irregular 
if the coupling is strengthened and trees with more generations are considered.  
For large $G$ and strong coupling firing 
stops [Fig.~\ref{stoch2.fig}(a1)] since  excitatory inputs to peripheral 
nodes are too weak to sustain firing of APs. In contrast, in oscillatory trees, 
the firing rate saturates for strong coupling [Fig.~\ref{stoch2.fig}(a2)] and firing becomes 
more regular if strongly-coupled trees with more generations are considered [Fig.~\ref{stoch2.fig}(b2)]. 
\begin{figure}[h]
\centering
\includegraphics[width=0.7\linewidth]{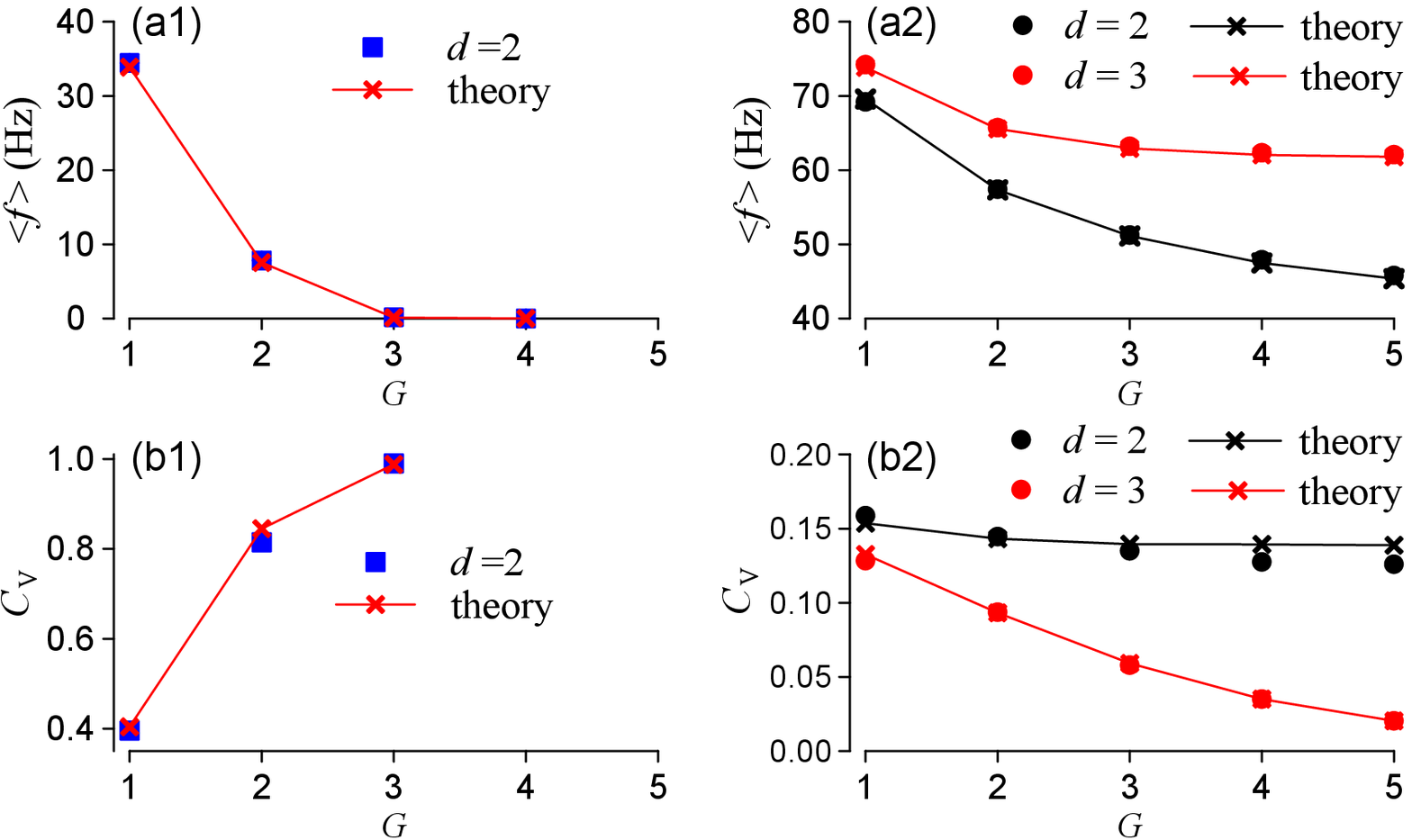}
\caption{Mean firing rate, $\mean{f}$, (a1, a2) and coefficient of variation, $C_V$,  of interspike intervals (b1, b2) of the central node versus the number of generations of tree networks 
in the strong coupling regime, 
$\kappa=1000$~mS/cm$^2$ and noise intensity $D=500$~\Dunits.  Left panels (a1, b1) correspond to an excitable tree with constant current 
$\constCurrent=20~\mu$A/cm$^2$; right panels (a2, b2) refer to an oscillatory tree with 
$\constCurrent=60~\mu$A/cm$^2$. Symbols $\square$ and $\bigcirc$ mark results of numerical simulations of the corresponding 
network with the indicated branching, $d$; solid lines and symbols $\times$ show theoretical scaling predictions. 
}
\label{stoch3.fig}
\end{figure}
\subsection*{Scaling of Effective Current and Noise intensity}
\label{sec:ScalingTheory}
In the strong coupling regime the dynamics of the central node of a tree network can be described by the dynamics of a single isolated node with membrane  potential $V_{0}(t)$ and 
with effective input current $\Ieff$ and an effective Gaussian noise with intensity $\Deff$, 
i.e. the influence of the coupling term on the dynamics can be approximated by a 
constant current and a white Gaussian noise. Then the dynamics of the membrane potential of the central node in 
eq. (\ref{model.eq2}) can be approximated by
\begin{eqnarray}
\label{eq:singleNodeDescription}
&&C \dot{V}_0 \approx f[V_0,\textbf{u}_0] + \Ieff + \sqrt{2 \Deff} \, \xi_{\text{eff}}(t),\\
&&\dot{\mathbf{u}}_0 \approx g[V_0,\textbf{u}_0]. \nonumber
\end{eqnarray}
%
%Here the functions $f[V_0,\textbf{u}_0]$ and $g[V_0,\textbf{u}_0]$ account for the intrinsic dynamics. In particular, 
%in case of the Hodgkin-Huxley-type node dynamics, we set $f[V_0,\textbf{u}_0]= - I_\text{ion} [V_0,\textbf{u}_0]$ and  $g[V_0,\textbf{u}_0]=\mathbf{\alpha}(V_0)(1-\mathbf{u}_0)-\mathbf{\beta}(V_0)(1-\mathbf{u}_0)$.
In the following, we derive those effective parameters for regular trees of diffusively-coupled nodes. 

In the network model, the dynamics of the membrane potential of $k$-th node is given by
\begin{eqnarray}
\label{eq:generalNodeDynamics}
C \dot{V}_k = f[V_k,\textbf{u}_k] 
+ \kappa \sum_{j=0}^{N-1} A_{k,j}(V_j - V_k) + I_{{\rm ext}_k}, \ \ \ k=0,1,...,N-1.
\end{eqnarray}
In order to derive approximations for the scaling of the effective current  $I_{{\rm ext}}$ and  the noise intensity $\Deff$, 
we extend the approach of Kouvaris et al. \cite{ko2014}, who considered the propagation of excitable waves in a 
tree network of identical Fitz-Hugh Nagumo nodes in the absence of noisy inputs.
Following their approach, we consider the dynamics of the average membrane potential 
$\langle V \rangle_g$ (termed density by Kouvaris et al.) in each shell in a tree.
Here and in the following $\langle  \rangle_g$ denotes averaging
over all nodes of the $g$th shell, 
\begin{eqnarray}
\langle V \rangle_g := \frac{1}{d^g} \sum \limits_{g\text{th shell}} V_k. 
\end{eqnarray} 

The dynamics of those densities can be obtained by averaging the respective equations for the dynamics of the membrane potentials, Eq. (\ref{eq:generalNodeDynamics}), over 
all nodes in one shell. Since the total number of connections between nodes in shell $g$ and $g-1$ is $d^g$, we obtain
\begin{eqnarray}
\label{eq:averageOverGenerationDynamics}
C\langle\dot{V} \rangle_g&=&\langle f(V,\textbf{u}) \rangle_g +
 \begin{cases}
 \kappa  d (\langle V \rangle_{1}-V_0),  & g=0,\\
 \kappa \left( \langle V \rangle_{g-1}-(d+1)\langle V \rangle_g+d \langle V \rangle_{g+1} \right), & 0<g<G, \\
 \kappa \left( \langle V \rangle_{G-1}-\langle V \rangle_G \right) + \constCurrent + \sqrt{2 \frac{D}{N_p}} \, \xi_G(t),  & g=G. \\
\end{cases}\notag\\
\end{eqnarray}
Note that, since peripheral nodes are subject to independent white Gaussian noises, 
the corresponding equation for the averaged membrane potentials of the peripheral generation contains white Gaussian noise
$\xi_G(t)$ with reduced intensity $\frac{D}{N_p}$. 

Since the coupling terms depend only on the difference between densities of the membrane potentials in 
adjacent generations $\Delta V_g := \langle V \rangle_g-\langle V \rangle_{g+1}$, we consider the dynamics of those 
differences next.
Subtracting equations for $\langle \dot{V} \rangle_g$ yields,
\begin{eqnarray}
C\Delta \dot{V}_g=\left( \langle f(V,\textbf{u})\rangle_g - \langle f(V,\textbf{u}) \rangle_{g+1}\right) + \Delta I_g + \Delta \xi_g(t)
+\begin{cases}
 \kappa \left( d \Delta V_{1}- (d+1) \Delta V_0 \right),  & g=0,\\
 & \\
 \kappa \left( \Delta V_{g-1}-(d+1) \Delta V_g+d \Delta V_{g+1} \right), & 0<g<G-1, \\
 & \\
 \kappa \left( \Delta V_{G-2}-  (d+1)\Delta V_{G-1} \right),  & g=G-1, \\
\end{cases}\notag \\
\end{eqnarray}
where $\Delta I_g = -\delta_{g,G-1}\, \constCurrent$ and $\Delta \xi_g(t) = -\delta_{g,G-1}\, \sqrt{2 \frac{D}{N_p}}\, \xi_G(t)$ is Gaussian white noise. 

Next, we consider the case of strong coupling. In that case, $\Delta V_{g}$ becomes small, and the membrane potentials of individual nodes approach 
the average potentials of the corresponding shell. Thus, we 
can approximate $\langle f(V,\textbf{u})\rangle_g - \langle f(V,\textbf{u}) \rangle_{g+1}$ by a Taylor expansion 
around  $\Delta V_g=0$, i.e. 
$\langle f(V,\textbf{u})\rangle_g - \langle f(V,\textbf{u}) \rangle_{g+1} \approx \Delta f_g^0 + \Delta f_g^1 \Delta V_g + h.o.$. 
It then follows for strong coupling, i.e. 
\begin{equation}
\label{eq:StrongCpl}
\Delta f_g^0=0, \quad \kappa \gg |\Delta f_g^1|$, $g=0,1,...,G-1,
\end{equation}
that the dynamics of the averaged potential is dominated by the coupling term and 
$\Delta V_g$ can be approximated by a multidimensional Ornstein$-$Uhlenbeck process,
\begin{eqnarray}
\label{eq:OUrepresentation}
C\frac{d}{dt}\boldsymbol{\Delta} \textbf{V}\approx \kappa \mathbf{B}  \boldsymbol{\Delta} \textbf{V} + \boldsymbol{\Delta} \textbf{I} + \boldsymbol{\Delta \xi}(t).
\end{eqnarray}
Here we introduced the $G$-dimensional vectors,
\begin{eqnarray}
\boldsymbol{\Delta} \textbf{V}=(\Delta V_0, \Delta V_1,..., \Delta V_{G-1})^T, \quad
\boldsymbol{\Delta} \textbf{I}=(\Delta \constCurrent,\Delta I_1,..., \Delta I_{G-1})^T, \quad
\boldsymbol{\Delta} \boldsymbol{\xi}(t)=(\Delta \xi_0(t),\Delta \xi_1(t),...,\Delta \xi_{G-1}(t))^T,\nonumber
\end{eqnarray}
and the $G \times G$ tridiagonal Toeplitz matrix,
\begin{eqnarray}
\label{eq:matrixBG} 
\mathbf{B}= \begin{pmatrix}
-\left(d+1\right) & d & 0 & ... & 0 \\
1 & -\left(d+1\right) & d & ... & ... \\
0 & 1 & -\left( d+1 \right) & ... & 0 \\
... & ... & ... & ... & d \\
0 & .. & 0 & 1 & -\left(d + 1 \right) \\
\end{pmatrix}.
\end{eqnarray} 
In the strong coupling limit (\ref{eq:StrongCpl}), deviations of $\boldsymbol{\Delta} \textbf{V}$ from its mean value decay extremely fast 
and we can use an adiabatic elimination \cite{vankampen1985elimination}  
to approximate $\boldsymbol{\Delta} \textbf{V}$ by its mean value plus a white Gaussian noise. Both, 
the mean voltage difference and the intensity of the Gaussian white noise in the strong coupling limit 
can be obtained by setting the left-hand side of Eq. (\ref{eq:OUrepresentation}) to zero. This yields 
\begin{equation}
\label{eq:DensityDifferences}
\boldsymbol{\Delta} \textbf{V} \approx - \frac{1}{\kappa}  \mathbf{B}^{-1} \left( \boldsymbol{\Delta} \textbf{I} + \boldsymbol{\Delta \xi}(t) \right),
\end{equation}
where $\mathbf{B}^{-1}$ is the inverse of the matrix $\mathbf{B}$. 
In order to obtain an approximation for the dynamics of the central node, we can use 
Eq. (\ref{eq:DensityDifferences}) to replace 
$V_{0}-\langle V \rangle_{1}$ by $\Delta V_{0}$ in Eq. (\ref{eq:averageOverGenerationDynamics}) 
for the central node, $g=0$. This yields
\begin{eqnarray}
\label{eq:singleNodeStep1}
C\langle\dot{V} \rangle_0=C \dot{V}_0=f[V_0,\textbf{u}_0] + d \mathbf{B}^{-1} \left( \boldsymbol{\Delta} \textbf{I} + \boldsymbol{\Delta \xi}(t) \right)_1. 
\end{eqnarray}
Here and in the following the index "$1$" denotes the first component of a $G$-dimensional vector. 
Next, the effective parameters $I_{\text{eff}}$ and $D_{\text{eff}}$ can be obtained by comparing   
Eqs. (\ref{eq:singleNodeStep1}) and (\ref{eq:singleNodeDescription}). 
This yields the effective input current and the intensity of the effective white Gaussian noise,
\begin{eqnarray}
\Ieff=d \left( \mathbf{B}^{-1} \boldsymbol{\Delta} \textbf{I} \right)_1, \quad 
\Deff=d \left( \mathbf{B}^{-1} \boldsymbol{\Delta \xi}(t) \right)_1. 
\end{eqnarray}

For the special case, considered in this study, 
that only peripheral nodes are subject to noisy inputs, i.e. 
 $\boldsymbol{\Delta} \textbf{I}=(0,0,..., -\constCurrent)^T$, and 
 $\boldsymbol{\Delta} \boldsymbol{\xi}(t)=(0,0,...,-\sqrt{2 D /N_p}\,\xi_{G}(t))^T$, 
 the calculation of the effective parameters $\Ieff$ and $\Deff$ requires only a single component, $(1,G)$, of
 the inverse matrix, $\mathbf{B}^{-1}$. Since $\mathbf{B}$ is a tridiagonal Toeplitz matrix, we 
can apply the results of Ref. \cite{da2001explicit} to calculate this component (see Supplemental Material for details on calculations) and find for the effective current, 
\begin{eqnarray}
\label{eq:effectiveCurrent}
\Ieff=\frac{N_p}{N} \constCurrent = R(G,d) \constCurrent, 
\end{eqnarray}
and for the effective noise intensity,
\begin{eqnarray}
\label{eq:effectiveNoiseCurrentToPeripherals}
\Deff=\frac{N_p}{N^2}\,D = \frac{R(G,d)}{N}\,D,
\end{eqnarray}
where the scaling factor $R(G,d)$ is given by Eq. (\ref{r.eq}).  
\begin{figure}[h!]
\centering
\includegraphics[width=0.7\linewidth]{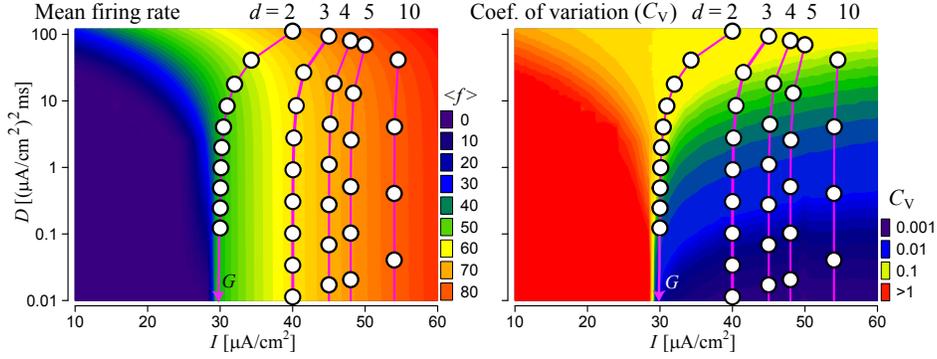}
\caption{Heat maps of the mean firing rate, $\mean{f}$ and coefficient of variation, $C_V$, vs 
input current, $\constCurrent$, and noise intensity, $D$, 
for the single  isolated HH node obtained from numerical simulations. Circles and magenta lines show the scaling of current, Eq. (\ref{eq:effectiveCurrent}), and 
noise intensity, Eq. (\ref{eq:effectiveNoiseCurrentToPeripherals}), 
for tree networks with indicated values of branching $d$ and with increasing number of generations, $G$  from 1 (top) to 10 (bottom). For tree networks the input 
current to the peripheral nodes is $\constCurrent=60~\mu$A/cm$^2$ and the noise intensity is $D=500$~\Dunits. 
}
\label{2Dscale.fig}
\end{figure}

Investigating the scaling of the effective parameters in more detail, we first note that our 
theory yields the scaling relation, Eq. (\ref{eq:scalingThresholdCurrent}), observed for 
the deterministic threshold current in Fig.~\ref{det2.fig}. In fact, the same scaling relation 
applies to the bifurcation values of $\constCurrent$ in the deterministic model, e.g. the 
subcritical Andronov-Hopf bifurcation of the equilibrium or the saddle-node bifurcation of 
the limit cycles.
Second, in Fig.~\ref{stoch3.fig}, we demonstrate the validity of the theoretical scaling 
predictions by comparing 
results for the mean firing rate and the CV from direct simulation of tree networks with those from a single 
node (\ref{eq:singleNodeDescription}) with input current and noise intensity scaled according to 
Eqs. (\ref{eq:effectiveCurrent}) and (\ref{eq:effectiveNoiseCurrentToPeripherals}), respectively. 
As illustrated in Fig.~\ref{2Dscale.fig}, we find an excellent correspondence of both results. 
This indicates that in the strong coupling limit the response of the network can be predicted from  
the stochastic dynamics of the effective central node. The statistics of 
interspike intervals for a single isolated node versus
input current parameters, i.e. constant component, $\constCurrent$, and noise intensity, $D$, can be easily computed numerically 
yielding two-dimensional maps, such as shown in Fig. \ref{2Dscale.fig}. 
Then for a given size (number of generations, $G$) and branching, $d$, of a tree, the scaled parameters, 
Eqs.~(\ref{eq:effectiveCurrent}) and (\ref{eq:effectiveNoiseCurrentToPeripherals}), set an operation point for 
the tree on 
the parametric map of a single element. Thus, predictions of the firing statistics of the central node of 
a tree of 
strongly-coupled excitable elements can be deduced by superimposing parametric dependencies of $I_\text{eff}$ 
and 
$D_\text{eff}$ on the parameters of the network. Figure~\ref{2Dscale.fig} demonstrates this for trees of 
strongly-coupled 
HH nodes in the oscillating regime. In trees with more generations $G$ the operation point 
is shifted towards smaller currents and lower noise intensities, resulting, for oscillatory trees, in slower 
and more coherent firing of the tree's central 
node. Finally, in the strong coupling limit the scaling relations (\ref{eq:effectiveCurrent}, 
\ref{eq:effectiveNoiseCurrentToPeripherals}) are independent of  the particular choice of the nodal model, e.g. they are expected to work 
for either Hodgkin-Huxley or Frankenhaeuser-Huxley nodal models.

\subsection*{Signal Detection}
The signal detection efficiency of a neuron can be quantified using the discriminability and the Fisher information \cite{seung1993,stemmler1996,brunel1998,pitkow2015,greenwood2016}.
In case of our model of coupled excitable elements on a tree, we use these measures to characterize how the tree topology affects its ability to distinguish between
two stimuli, $\constCurrent$ and $\constCurrent +\Delta \constCurrent$, applied to the peripheral nodes.

The preceding section showed that in the strong coupling limit, the stochastic dynamics of the network 
could be predicted from the dynamics of a single node with appropriately scaled parameters of the 
input current. Thus, we first analyze the lower bound of the Fisher information 
of a single node. Equation (\ref{fisher.eq}) indicates that  the Fisher information is determined by 
two factors: the term $\left(\frac{d \mu_T}{d\constCurrent}\right)^2$, which is related to the slope of the so-called $f-I$ 
curve (mean firing rate vs input current curve) 
 and determines the sensitivity of a neuron to small variations of the input current. 
The sensitivity is largest in the vicinity of the bifurcation point, where the limit cycle is born, 
and where the slope of the $f-I$ curve is the steepest. In this region, the Fisher information is high. 
However, the second factor in Eq. (\ref{fisher.eq}), the variance of the spike count,  may degrade the Fisher information. 
In the excitable regime, when the input current is below its bifurcation value and APs are 
induced by noise, the phenomenon of stochastic resonance is observed \cite{gammaitoni1998stochastic}, i.e. due to 
the competition of two factors, the sensitivity and the spike count variance, the Fisher information possesses a maximum at an optimal noise intensity \cite{stemmler1996}. 

\begin{figure} [h!] 
  \centering
   \includegraphics[width=0.7\textwidth]{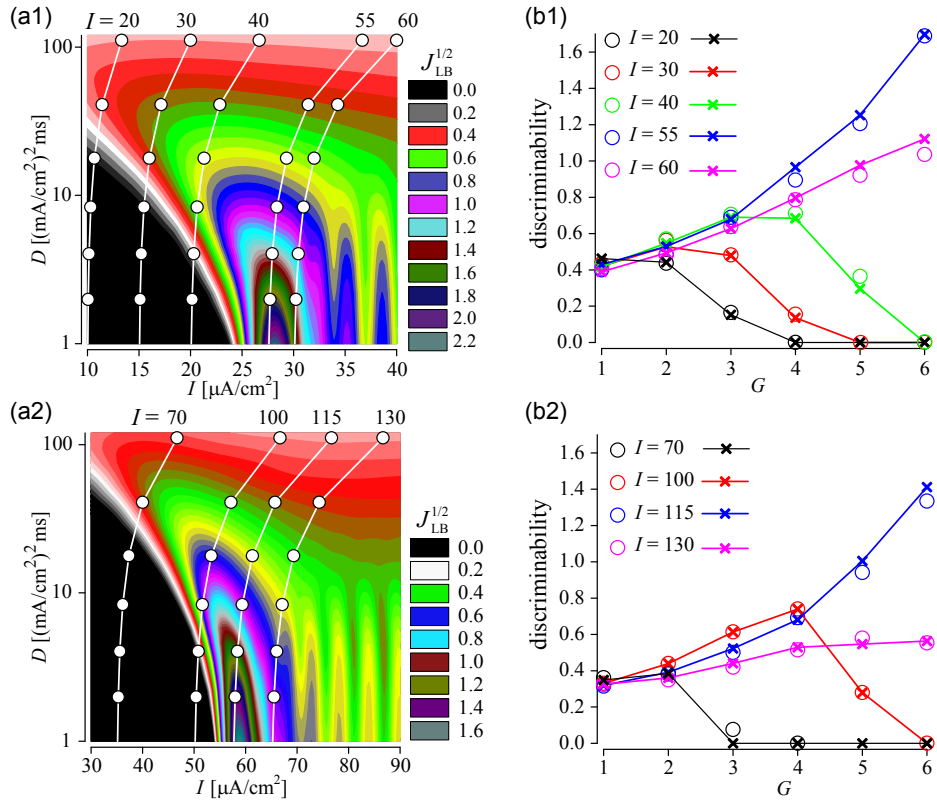}
   \caption{Signal detection by a single node and by tree networks. 
   (a): Square root of the lower bound of the Fisher information (\ref{fisher.eq}) 
   of the single HH (a1) and single FH (a2) node versus input current and noise intensity. 
   Superimposed white lines with circle symbols show the scaling of the effective input current, Eq. (\ref{eq:effectiveCurrent}), and effective noise 
   intensity, Eq. (\ref{eq:effectiveNoiseCurrentToPeripherals}), for tree networks with branching ratio $d=2$ 
   and increasing number of generations. Each of these lines corresponds to values of the external current to peripheral nodes, $\constCurrent$, 
   indicated at the top of panels (a). The number of generations, $G$, increases from 1 (top) to 6 (bottom). Noise intensity for tree networks is $D=500$~[\Dunits].  
   (b): Discriminability  (\ref{discrim.eq1}) versus the number of generations, $G$, for tree networks 
   with branching ratio $d=2$ and indicated 
   values of the input current $\constCurrent$ for HH (b1) and FH (b2) nodes. The increment of the input current is $\Delta \constCurrent = 2$~$\mu$A/cm$^2$ and noise intensity is $D=500$~[\Dunits]. Filled 
   circles refer to results of numerical simulations of corresponding tree networks; solid lines and symbols $\times$ show 
   theoretical scaling predictions obtained from simulations of the single HH or FH node with the input current and noise intensity scaled according to 
   Eq. (\ref{eq:effectiveCurrent}) and Eq. (\ref{eq:effectiveNoiseCurrentToPeripherals}), respectively.
In all panels the spike count statistics was calculated for $T=200$~ms windows. The coupling strength 
on the panels (b1, b2) is $\kappa = 1000$~mS/cm$^2$. 
   }
\label{fisher.fig}
\end{figure}
Figure~\ref{fisher.fig}(a1) shows the lower bound of the Fisher information, $J_\text{LB}$, for a single 
HH node as a function of input current and noise intensity.  
The Fisher information is maximal for an input current $\constCurrent$, which brings the system close to the 
transition to periodic spiking, i.e.  28 -- 29~$\mu$A/cm$^2$.  In Fig.~\ref{fisher.fig}(a1), a 
vertical section across the  map corresponds to the dependence of the Fisher information on noise 
intensity. As can be seen, such a dependence is 
non-monotonous in the excitable regimes, e.g. for $\constCurrent=15$ or $\constCurrent=20$~$\mu$A/cm$^2$, indicating 
the phenomenon of stochastic resonance \cite{gammaitoni1998stochastic}, reported 
before for the original Hodgkin-Huxley neuron model in Ref.\cite{stemmler1996}. Indeed, stochastic 
resonance is a generic phenomenon in excitable systems \cite{gammaitoni1998stochastic,lin2000} and 
so the Frankenhaeuser-Huxley (FH) nodal model demonstrates qualitatively similar parameter dependence, 
shown in Fig.~\ref{fisher.fig}(a2). In the absence of noise, a stable equilibrium of the single FH node passes through 
a subcritical Andronov-Hopf bifurcation at $\IAH \approx 60.19$~$\mu$A/cm$^2$. Consequently, the Fisher information in 
Fig.~\ref{fisher.fig}(a2) is maximal around this value, similar to the HH node. In the excitable regime, 
e.g. $\constCurrent=40$~~$\mu$A/cm$^2$, the Fisher information vs. noise intensity passes through a maximum, 
demonstrating stochastic resonance, again, qualitatively similar to the HH node.

The scaling relations for the input current, Eq. (\ref{eq:effectiveCurrent}), and noise intensity, 
Eq. (\ref{eq:effectiveNoiseCurrentToPeripherals}), enable us to 
predict the signal detection capability of a tree network in the regime of strong coupling. Given 
the branching, $d$ and the input current to the 
peripheral nodes, an increase in the number of generations (i.e. tree size) results in a 
decrease of the effective input current and noise. Then, depending on the particular values of 
$\constCurrent$ and $D$, signal detection by the tree may show distinct dependencies on the tree size, $G$. 
This is illustrated in Fig.~\ref{fisher.fig}(a1,a2) by superimposing the scaling of the input 
current and noise intensity on the Fisher information 
map of the single node. In particular, our theory predicts that in the excitable regime, i.e. when the 
network does not produce sustained 
periodic firing in the absence of stochastic inputs, the scaling of $\constCurrent$ and $D$  may bring an effective operating point of the network 
across the local maximum of the Fisher information.
As can be seen, for instance, for the input current $\constCurrent=30$ or $40$~$\mu$A/cm$^2$ for HH nodal model, 
an increase of the number of 
generations to $G=$2--4 brings the effective operating point to regions of higher values of the Fisher information; further growth of 
the tree size eventually suppresses AP firing and thus small signals cannot be detected. 
In contrast, in the oscillatory regime 
(e.g. for $\constCurrent=60$~$\mu$A/cm$^2$ for HH nodal model), the increase of the network size moves 
the operation point always to regions of higher values of 
the Fisher information and so the discriminability increases monotonously with the tree size, $G$. 
Interestingly, one could predict the 
input current to the network which for a tree with large enough generations would result in an 
effective operating point close to bifurcation 
value of the single node. For example, for the HH nodes, such a value of the external current is 
$\constCurrent=55$~$\mu$A/cm$^2$ and for the FH node, $\constCurrent=115$~$\mu$A/cm$^2$. For such currents increasing 
the tree size should result in a higher degree of signal discrimination. 

To test these predictions we computed the discriminability (\ref{discrim.eq1}) for trees with different 
numbers of generations and for the single node with the scaled values of constant input current and 
noise intensity according to Eqs.~(\ref{eq:effectiveCurrent}, \ref{eq:effectiveNoiseCurrentToPeripherals}). 
 Figure~\ref{fisher.fig}(b1,b2) shows excellent correspondence between the respective discriminabilities.

\section*{Conclusion}
We have studied the emergence of noisy periodic spiking in regular tree networks 
of coupled  excitable elements. Using biophysical models of excitable nodes, 
we showed that noisy periodic network spiking can be generated, although the 
periphery of the tree is excited by random and independent inputs (Fig.~\ref{stoch1.fig}). The firing rate and coherence of spiking
can be maximized by varying the coupling strength and is altered by changing the network topology (Figs.~\ref{stoch2.fig},\ref{stoch3.fig}). 

We put special emphasis on the strong coupling regime, which refers to the case 
of excitable nodes of Ranvier linked by myelinated (dendrite or axon) fibers of a neuron.

It is intuitively clear that in the strong coupling limit, the collective 
dynamics of the network could be described by a single effective excitable system. 
We have derived the corresponding scaling relations for random inputs Eqs. 
(\ref{eq:effectiveCurrent}, \ref{eq:effectiveNoiseCurrentToPeripherals}) which allows for reliable 
predictions of the collective network response 
based on the stochastic dynamics of a single isolated node with 
scaled input parameters. Stochastic excitable systems demonstrate non-trivial behaviour 
versus the noise intensity. Examples include the phenomena of coherence resonance 
\cite{PRLpikovsky97}, whereby the variability of spiking events 
(e.g. coefficient of variation) is minimal for non-zero noise intensity, and 
stochastic resonance, characterized by non-monotonous dependence of a response to 
an external signal on the noise intensity \cite{gammaitoni1998stochastic}. 
Similar phenomena have been observed in networks of excitable elements. In particular, 
the phenomena of system size stochastic \cite{Pikovsky2002} and coherence resonance 
\cite{toral2003}, which are also observed in strongly-coupled star networks of excitable 
elements \cite{justus2016}. As we have shown in the present paper, the phenomenon of system size stochastic resonance 
also occurs in strongly coupled tree networks, i.e. the number of generations in a tree network of 
excitable elements can be tuned in order to optimize the network ability to discriminate 
between different input signals. In particular, our analytical approach allows for the prediction 
of optimal tree sizes and branching ratio. 

The analytical approach developed here can be extended to random 
trees \cite{drmota2009random} in which the branching ratio varies among different generations, 
yielding similar scaling relations in the strong coupling limit. While we considered 
networks of identical nodes, our approach can be readily extended to the inhomogeneous case, as long as the condition for strong coupling Eq. (\ref{eq:StrongCpl}) is satisfied.

Our results suggest a mechanism for the emergence of noisy periodic firing  and information coding by peripheral sensory neurons which possess branched tree-like myelinated dendrites \cite{banks1997pacemaker}. Such neurons may possess multiple spike initiation zones at peripheral nodes (heminodes) and nodes of Ranvier at branching points. 
Examples of the muscle spindles \cite{banks1982form} and cutaneous  mechanoreceptors \cite{lesniak2014computation} indicate that myelinated dendritic trees extend to up to 7 generations.  
Myelin provides low-resistance links between nodes and fast saltatory conduction of APs, 
which corresponds to strong coupling between the nodes of Ranvier. For example, the average diameter of a cat muscle spindles afferents ranges 
from $3$ to $13$~$\mu$m, while links between nodes are relatively short, 
$50$ -- $200$~$\mu$m \cite{banks1982form}. An estimate of the coupling strength from Eq. (\ref{model.eq3}) yields values well 
within the range of the strong coupling regime used in our study. 
The collective noisy periodic firing then may occur due to the synchronized noise-induced generation of APs by stimulating the peripheral heminodes, as described by our model.
Given the biophysical properties of the nodes of 
Ranvier and the sensory inputs, the variability of interspike intervals and 
the stimulus discrimination capability of a neuron are determined by the ratio of 
the number of signal-receiving peripheral heminodes to the total number of 
nodes in the network.

\section*{Acknowledgements}
We thank D.F.~Russell, E.~Schoell, T.~Isele for fruitful discussions. AN acknowledges support by the Lobachevsky University of Nizhny Novgorod through the Russian Science Foundation grant 14-41-000440. LSG thanks Ohio University for hospitality and support.

\bigskip
\noindent
{\bf Author contributions.}
AN formulated the problem. JK performed analytical calculation.
AKN and AN performed numerical simulations.
JK, AKN, LSG and AN  wrote and reviewed the manuscript.

\bigskip
\noindent
{\bf Competing financial interests.}
The authors declare no competing financial interests.

\newpage
\flushbottom

\section*{Supplementary material.\\
Emergent stochastic oscillations and signal detection in regular tree networks of strongly coupled excitable elements}

\section*{Justus A. Kromer, Ali Khaledi-Nasab, Lutz Schimansky-Geier, Alexander B. Neiman}
\bigskip

\subsection*{Models for Nodes of Ranvier}
\appendix
A Hodgkin-Huxley type model (HH) for a  node of Ranvier contains only sodium and leak ionic currents. 
Thus, in eq.(2) of the main paper, the ionic current becomes $I_\text{ion} = I_\text{Na}+I_\text{L}$. For the sodium current we used the 
Hodgkin-Huxley (HH) type kinetics \cite{mc2002,hh},
$I_\text{Na} = g_\text{Na} m^3 h (V-V_\text{Na})$, 
where $g_\text{Na} = 1100$~mS/cm$^2$ is the maximal value of the sodium conductance and $V_\text{Na}= 50$~mV is the Na reversal potential.
The gating activation and inactivation variables obey the dynamics 
\begin{eqnarray}
  \label{e3}
    &\dot{m}& = \alpha_m(V) (1-m)-\beta_m(V)m \cr
    &\dot{h}_k&  = \alpha_h (V)(1-h)-\beta_h(V)h,
\end{eqnarray}
with the following rate functions:
\begin{eqnarray}
  \label{eq4}
   &\alpha_m(V) &= 1.314(V+20.4)/[1-\exp[-(V+20.4)/10.3]],\cr
   &\beta_m(V)  &= -0.0608(V+25.7)/[1-\exp[(V+25.7)/11]], \cr
   &\alpha_h(V) &= -0.068(V+114)/[1-\exp[(V+114)/11]], \cr
   &\beta_h(V)  &= 2.52/[1+\exp[-(V+31.8)/13.4]].      
\end{eqnarray}
The leak current is $I_\text{L}=g_\text{L}(V_k-V_\text{L})$ with $g_\text{L}=20$~mS/cm$^2$ and 
$V_\text{L}=-80$~mV.

\bigskip

The Frankenhaeuser-Huxley (FH) model \cite{fh1964} uses four ionic currents: sodium, potassium, persistent sodium and leak:  $I_\text{ion} = I_\text{Na}+I_\text{K}+I_\text{p}+I_\text{L}$.
The currents are given by
 \begin{eqnarray}
 &I_{X}&= P_{X} \frac{EF^{2}}{RT} \frac{[X]_{0}-[X]_{i} e^{\frac{EF}{RT}}}{1-e^{\frac{EF}{RT}}}, \quad X=\text{Na, K, p}, \cr
 &I_\text{L}& = g_\text{L}(V-V_\text{L}),
 \end{eqnarray}
where $E=-70+V$; $P_X$ are the permeabilities of sodium (Na), potassium (K) and persistent (p) ionic currents:
 \begin{equation}
 P_\text{Na}=\bar{P}_\text{Na} h m^2, \quad
 P_\text{K}=\bar{P}_\text{K} n^2, \quad
 P_\text{p}=\bar{P}_\text{p} p^2,
 \label{e1}
\end{equation}
where $\bar{P}_X$ are maximal values of permeabilities of corresponding channels. In Eq. 
(\ref{e1}) $[X]_0$ and $[X]_i$ are corresponding extracellular and intracellular ionic concentrations, 
of Na and K ions, respectively; for the persistent current, $X=p$, we have $[X]\equiv[\text{Na}]$.
The gating variables follow,
\begin{eqnarray}
    &\dot{m}& = [\alpha_m(V) (1-m)-\beta_m(V)m]s, \cr
    &\dot{h}&  = [\alpha_h (V)(1-h)-\beta_h(V)h]s,\cr
        &\dot{n}& = [\alpha_n(V) (1-n)-\beta_n(V)n]s, \cr
    &\dot{p}&  = [\alpha_p (V)(1-p)-\beta_p(V)p]s,
\end{eqnarray}
 and the rate functions are:
\begin{eqnarray}
\alpha_m&=& 0.36(V-22)/(1- \exp[(22-V)/3]), \cr
\beta_m&=& 0.4 (13-V)/(1- \exp[(V-13)/20]), \cr
\alpha_h&=& 0.1 (10+V)/(\exp[(V+10)/6]-1),   \cr
\beta_h&=& 4.5/(1+ \exp[(45-V)/10]), \cr
\alpha_p&=&  0.006 (V-40)/(1-\exp[(40-V)/10]),  \cr
\beta_p&=&   0.09 (25+V)/(\exp[(V+25)/20]-1),  \cr
\alpha_n&=&   0.02 (V-35)/(1-\exp[(35-V)/10]), \cr
\beta_n&=&   0.05 (10-V)/(1-\exp[(V-10)/10]), \nonumber
\end{eqnarray}
where $s$ is a scaling factor.
The parameters for the model were taken from the original Frankenhaeuser-Huxley paper \cite{fh1964} 
with three modifications to reduce the frequency of periodic spiking when a  sufficiently-high   
constant current is injected: (i) the rate equations for the gating variables included a scale factor $s=0.3$; (ii) the maximal permeability of potassium channels is reduced to
$\bar{P}_\text{K}= 3.6\times 10^{-4}$~cm/sec;  and (iii) the leak conductance was reduced to 
$g_\text{L}=6$~mS/cm$^2$. 
Other parameters are the same as in the original FH model: 
$[Na]_{0}=114.5$~mM, $[Na]_{i}=13.74$~mM, $[K]_{0} =2.5$~mM, $[K]_{i} =120$~mM; $\bar{P}_\text{Na}= 8\times 10^{-3}$~cm/sec; 
$\bar{P}_\text{p}= 54\times 10^{-5}$~cm/sec; $V_\text{L}=0.026$~mV. Constants $R$ and $F$ are the universal gas constant and the Faraday constant, and $T=293.15$~K.

\subsection*{Inverse of Tridiagonal Toeplitz Matrix}
\label{sub:InverseToeplitzMatrix}
In order to calculate the effective current $I_{\text{mod}}$ eq. (20) and noise intensity 
$D_{\text{mod}}$ eq. (21), 
we need to evaluate the inverse matrix $\mathbf{B}^{-1}$ of $\mathbf{B}$. To this end, 
we apply results from Ref. \cite{da2001explicit} in which the components of the inverse of a tridiagonal Toeplitz matrix are given in terms of 
two sequences $\{v_i\}$ and $\{u_i\}$ with $i=1,2,...,G$. If we apply their results 
to the matrix $\mathbf{B}$, eq. (17), we obtain for its $(i,j)$th component
\begin{align}
\label{eq:ComponentsInverseTridiagonalMatrix}
\mathbf{B}^{-1}_{ij} =  u_{\text{Min(i,j)}} v_{\text{Max(i,j)}} 
\begin{cases}
\frac{d^i}{d^j} & \text{if } i<j\\
1 & \text{else} 
\end{cases}.
\end{align}
The sequences $\{u_i\}$ and $\{v_i\}$ are only defined up to a multiplicative constant and it is convenient to set $u_1=1$. Then the 
sequences can be calculated from 
\begin{eqnarray*}
\label{eq:vsequence}
v_1 = \frac{1}{d_1}, \ v_k = -\frac{d}{d_k} v_{k-1}, \ \ \ k=2,3,..,G
\end{eqnarray*}
with
\begin{eqnarray*}
\label{eq:dsequence}
d_{G} = -\left(d+1\right), \ d_i = -\left(d+1\right) - \frac{d}{d_{i+1}}, \ \ \ i=1, ..., G-2,G-1
\end{eqnarray*}
and
\begin{eqnarray*}
u_{G} = \frac{1}{w_{G} v_{G}}, \ u_k = -\frac{d}{w_k} u_{k+1}, \ \ \ k=1, ...., G-2,G-1
\end{eqnarray*}
with
\begin{eqnarray*}
w_1 = -(n_1+1), \ w_i = -\left(d+1\right) - \frac{d}{w_{i-1}}, \ \ \ i=2,3,..,G.
\end{eqnarray*}

Since we are interested in the dynamics of the first order node ($g=0$) when only peripherals are 
subject to noisy currents, we only need to evaluate a single component, $(1,G)$,  of $\mathbf{B}^{-1}$. For this component we find from eq. (\ref{eq:ComponentsInverseTridiagonalMatrix}), 
$\mathbf{B}^{-1}_{1,G} = v_G$. Applying the definition of $v_k$, eq. (\ref{eq:vsequence}), multiple times, we 
obtain 
\begin{eqnarray}
v_G=(-1)^{G-1} \frac{d^{G-1}}{\prod \limits_{k=1}^{G}d_k}.
\end{eqnarray} 
Next, we evaluate the product for which one can show by using mathematical induction that 
\begin{eqnarray}
\prod \limits_{k=1}^{j} d_k= (-1)^j \left( \sum \limits_{k=0}^{j} d^k - \frac{d}{d_{j+1}} \sum \limits_{l=0}^{j-1} d^k \right), \ \ j<G.
\end{eqnarray}
Finally, multiplication by $d_G$ and applying eq. (\ref{eq:dsequence}) yields 
\begin{eqnarray}
\prod \limits_{k=1}^{G} d_k=(-1)^{G} \sum \limits_{l=0}^{G} d^l = (-1)^{G} N.
\end{eqnarray}
This yields for the $(1,G)$ component of $\mathbf{B}^{-1}$
\begin{eqnarray}
\mathbf{B}^{-1}_{1G}=- \frac{d^{G-1}}{N}.
\end{eqnarray}
Using this in eqs. (19) yields the effective current and noise intensity, respectively.


\begin{thebibliography}{10}
\expandafter\ifx\csname url\endcsname\relax
  \def\url#1{\texttt{#1}}\fi
\expandafter\ifx\csname urlprefix\endcsname\relax\def\urlprefix{URL }\fi
\providecommand{\bibinfo}[2]{#2}
\providecommand{\eprint}[2][]{\url{#2}}

\bibitem{kiss2003chemical}
\bibinfo{author}{Kiss, I.~Z.} \& \bibinfo{author}{Hudson, J.~L.}
\newblock \bibinfo{title}{Chemical complexity: Spontaneous and engineered
  structures}.
\newblock \emph{\bibinfo{journal}{AIChE journal}}
  \textbf{\bibinfo{volume}{49}}, \bibinfo{pages}{2234--2241}
  (\bibinfo{year}{2003}).

\bibitem{mikhailov2012engineering}
\bibinfo{author}{Mikhailov, A.~S.} \& \bibinfo{author}{Ertl, G.}
\newblock \emph{\bibinfo{title}{Engineering of Chemical Complexity}}
  (\bibinfo{publisher}{World Scientific}, \bibinfo{year}{2012}).

\bibitem{farkas2002social}
\bibinfo{author}{Farkas, I.}, \bibinfo{author}{Helbing, D.} \&
  \bibinfo{author}{Vicsek, T.}
\newblock \bibinfo{title}{Social behaviour: Mexican waves in an excitable
  medium}.
\newblock \emph{\bibinfo{journal}{Nature}} \textbf{\bibinfo{volume}{419}},
  \bibinfo{pages}{131--132} (\bibinfo{year}{2002}).

\bibitem{newman2002spread}
\bibinfo{author}{Newman, M.~E.}
\newblock \bibinfo{title}{Spread of epidemic disease on networks}.
\newblock \emph{\bibinfo{journal}{Physical Review E}}
  \textbf{\bibinfo{volume}{66}}, \bibinfo{pages}{016128}
  (\bibinfo{year}{2002}).

\bibitem{perc2006coherence}
\bibinfo{author}{Perc, M.}
\newblock \bibinfo{title}{Coherence resonance in a spatial prisoner's dilemma
  game}.
\newblock \emph{\bibinfo{journal}{New Journal of Physics}}
  \textbf{\bibinfo{volume}{8}}, \bibinfo{pages}{22} (\bibinfo{year}{2006}).

\bibitem{borgatti2009network}
\bibinfo{author}{Borgatti, S.~P.}, \bibinfo{author}{Mehra, A.},
  \bibinfo{author}{Brass, D.~J.} \& \bibinfo{author}{Labianca, G.}
\newblock \bibinfo{title}{Network analysis in the social sciences}.
\newblock \emph{\bibinfo{journal}{Science}} \textbf{\bibinfo{volume}{323}},
  \bibinfo{pages}{892--895} (\bibinfo{year}{2009}).

\bibitem{chen2015emergent}
\bibinfo{author}{Chen, Y.}, \bibinfo{author}{Kim, J.~K.},
  \bibinfo{author}{Hirning, A.~J.}, \bibinfo{author}{Josi{\'c}, K.} \&
  \bibinfo{author}{Bennett, M.~R.}
\newblock \bibinfo{title}{Emergent genetic oscillations in a synthetic
  microbial consortium}.
\newblock \emph{\bibinfo{journal}{Science}} \textbf{\bibinfo{volume}{349}},
  \bibinfo{pages}{986--989} (\bibinfo{year}{2015}).

\bibitem{Copelli2005691}
\bibinfo{author}{Copelli, M.}, \bibinfo{author}{Oliveira, R.~F.},
  \bibinfo{author}{Roque, A.~C.} \& \bibinfo{author}{Kinouchi, O.}
\newblock \bibinfo{title}{Signal compression in the sensory periphery}.
\newblock \emph{\bibinfo{journal}{Neurocomputing}}
  \textbf{\bibinfo{volume}{65–66}}, \bibinfo{pages}{691 -- 696}
  (\bibinfo{year}{2005}).

\bibitem{furtado2006response}
\bibinfo{author}{Furtado, L.~S.} \& \bibinfo{author}{Copelli, M.}
\newblock \bibinfo{title}{Response of electrically coupled spiking neurons: a
  cellular automaton approach}.
\newblock \emph{\bibinfo{journal}{Physical Review E}}
  \textbf{\bibinfo{volume}{73}}, \bibinfo{pages}{011907}
  (\bibinfo{year}{2006}).

\bibitem{maxim2015}
\bibinfo{author}{Moldakarimov, S.}, \bibinfo{author}{Bazhenov, M.} \&
  \bibinfo{author}{Sejnowski, T.~J.}
\newblock \bibinfo{title}{Feedback stabilizes propagation of synchronous
  spiking in cortical neural networks}.
\newblock \emph{\bibinfo{journal}{Proceedings of the National Academy of
  Sciences}} \textbf{\bibinfo{volume}{112}}, \bibinfo{pages}{2545--2550}
  (\bibinfo{year}{2015}).

\bibitem{jung1995spatiotemporal}
\bibinfo{author}{Jung, P.} \& \bibinfo{author}{Mayer-Kress, G.}
\newblock \bibinfo{title}{Spatiotemporal stochastic resonance in excitable
  media}.
\newblock \emph{\bibinfo{journal}{Physical Review Letters}}
  \textbf{\bibinfo{volume}{74}}, \bibinfo{pages}{2130} (\bibinfo{year}{1995}).

\bibitem{lin2000}
\bibinfo{author}{Lindner, B.}, \bibinfo{author}{Garc{\i}a-Ojalvo, J.},
  \bibinfo{author}{Neiman, A.} \& \bibinfo{author}{Schimansky-Geier, L.}
\newblock \bibinfo{title}{Effects of noise in excitable systems}.
\newblock \emph{\bibinfo{journal}{Physics Reports}}
  \textbf{\bibinfo{volume}{392}}, \bibinfo{pages}{321--424}
  (\bibinfo{year}{2004}).

\bibitem{sagues2007spatiotemporal}
\bibinfo{author}{Sagu{\'e}s, F.}, \bibinfo{author}{Sancho, J.~M.} \&
  \bibinfo{author}{Garc{\'\i}a-Ojalvo, J.}
\newblock \bibinfo{title}{Spatiotemporal order out of noise}.
\newblock \emph{\bibinfo{journal}{Reviews of Modern Physics}}
  \textbf{\bibinfo{volume}{79}}, \bibinfo{pages}{829} (\bibinfo{year}{2007}).

\bibitem{toral2003}
\bibinfo{author}{Toral, R.}, \bibinfo{author}{Mirasso, C.~R.} \&
  \bibinfo{author}{Gunton, J.~D.}
\newblock \bibinfo{title}{System size coherence resonance in coupled
  fitzhugh-nagumo models}.
\newblock \emph{\bibinfo{journal}{EPL (Europhysics Letters)}}
  \textbf{\bibinfo{volume}{61}}, \bibinfo{pages}{162} (\bibinfo{year}{2003}).

\bibitem{perc2005spatial}
\bibinfo{author}{Perc, M.}
\newblock \bibinfo{title}{Spatial coherence resonance in excitable media}.
\newblock \emph{\bibinfo{journal}{Physical Review E}}
  \textbf{\bibinfo{volume}{72}}, \bibinfo{pages}{016207}
  (\bibinfo{year}{2005}).

\bibitem{perc2007stochastic}
\bibinfo{author}{Perc, M.}
\newblock \bibinfo{title}{Stochastic resonance on excitable small-world
  networks via a pacemaker}.
\newblock \emph{\bibinfo{journal}{Physical Review E}}
  \textbf{\bibinfo{volume}{76}}, \bibinfo{pages}{066203}
  (\bibinfo{year}{2007}).

\bibitem{pregosak2010}
\bibinfo{author}{Gosak, M.}, \bibinfo{author}{Koro\ifmmode~\check{s}\else
  \v{s}\fi{}ak, D.} \& \bibinfo{author}{Marhl, M.}
\newblock \bibinfo{title}{Optimal network configuration for maximal coherence
  resonance in excitable systems}.
\newblock \emph{\bibinfo{journal}{Physical Review E}}
  \textbf{\bibinfo{volume}{81}}, \bibinfo{pages}{056104}
  (\bibinfo{year}{2010}).

\bibitem{kal2010}
\bibinfo{author}{Kaluza, P.}, \bibinfo{author}{Strege, C.} \&
  \bibinfo{author}{Meyer-Ortmanns, H.}
\newblock \bibinfo{title}{Noise as control parameter in networks of excitable
  media: role of the network topology}.
\newblock \emph{\bibinfo{journal}{Physical Review E}}
  \textbf{\bibinfo{volume}{82}}, \bibinfo{pages}{036104}
  (\bibinfo{year}{2010}).

\bibitem{son2013}
\bibinfo{author}{Sonnenschein, B.}, \bibinfo{author}{Zaks, M.},
  \bibinfo{author}{Neiman, A.} \& \bibinfo{author}{Schimansky-Geier, L.}
\newblock \bibinfo{title}{Excitable elements controlled by noise and network
  structure}.
\newblock \emph{\bibinfo{journal}{The European Physical Journal Special
  Topics}} \textbf{\bibinfo{volume}{222}}, \bibinfo{pages}{2517--2529}
  (\bibinfo{year}{2013}).

\bibitem{gollo2013single}
\bibinfo{author}{Gollo, L.~L.}, \bibinfo{author}{Kinouchi, O.} \&
  \bibinfo{author}{Copelli, M.}
\newblock \bibinfo{title}{Single-neuron criticality optimizes analog dendritic
  computation}.
\newblock \emph{\bibinfo{journal}{Scientific Reports}}
  \textbf{\bibinfo{volume}{3}} (\bibinfo{year}{2013}).

\bibitem{kinouchi2006optimal}
\bibinfo{author}{Kinouchi, O.} \& \bibinfo{author}{Copelli, M.}
\newblock \bibinfo{title}{Optimal dynamical range of excitable networks at
  criticality}.
\newblock \emph{\bibinfo{journal}{Nature Physics}}
  \textbf{\bibinfo{volume}{2}}, \bibinfo{pages}{348--351}
  (\bibinfo{year}{2006}).

\bibitem{larremore2011predicting}
\bibinfo{author}{Larremore, D.~B.}, \bibinfo{author}{Shew, W.~L.} \&
  \bibinfo{author}{Restrepo, J.~G.}
\newblock \bibinfo{title}{Predicting criticality and dynamic range in complex
  networks: effects of topology}.
\newblock \emph{\bibinfo{journal}{Physical Review Letters}}
  \textbf{\bibinfo{volume}{106}}, \bibinfo{pages}{058101}
  (\bibinfo{year}{2011}).

\bibitem{london2005dendritic}
\bibinfo{author}{London, M.} \& \bibinfo{author}{H{\"a}usser, M.}
\newblock \bibinfo{title}{Dendritic computation}.
\newblock \emph{\bibinfo{journal}{Annu. Rev. Neurosci.}}
  \textbf{\bibinfo{volume}{28}}, \bibinfo{pages}{503--532}
  (\bibinfo{year}{2005}).

\bibitem{st2016}
\bibinfo{author}{Stuart, G.}, \bibinfo{author}{Spruston, N.} \&
  \bibinfo{author}{H{\"a}usser, M.}
\newblock \emph{\bibinfo{title}{Dendrites}} (\bibinfo{publisher}{Oxford
  University Press}, \bibinfo{year}{2016}).

\bibitem{eagles1974afferent}
\bibinfo{author}{Eagles, J.~P.} \& \bibinfo{author}{Purple, R.~L.}
\newblock \bibinfo{title}{Afferent fibers with multiple encoding sites}.
\newblock \emph{\bibinfo{journal}{Brain Research}}
  \textbf{\bibinfo{volume}{77}}, \bibinfo{pages}{187--193}
  (\bibinfo{year}{1974}).

\bibitem{quick1980anatomical}
\bibinfo{author}{Quick, D.}, \bibinfo{author}{Kennedy, W.} \&
  \bibinfo{author}{Poppele, R.}
\newblock \bibinfo{title}{Anatomical evidence for multiple sources of action
  potentials in the afferent fibers of muscle spindles}.
\newblock \emph{\bibinfo{journal}{Neuroscience}} \textbf{\bibinfo{volume}{5}},
  \bibinfo{pages}{109--115} (\bibinfo{year}{1980}).

\bibitem{banks1982form}
\bibinfo{author}{Banks, R.~W.}, \bibinfo{author}{Barker, D.} \&
  \bibinfo{author}{Stacey, M.}
\newblock \bibinfo{title}{Form and distribution of sensory terminals in cat
  hindlimb muscle spindles}.
\newblock \emph{\bibinfo{journal}{Philosophical Transactions of the Royal
  Society of London B: Biological Sciences}} \textbf{\bibinfo{volume}{299}},
  \bibinfo{pages}{329--364} (\bibinfo{year}{1982}).

\bibitem{banks1997pacemaker}
\bibinfo{author}{Banks, R.}, \bibinfo{author}{Hulliger, M.},
  \bibinfo{author}{Scheepstra, K.} \& \bibinfo{author}{Otten, E.}
\newblock \bibinfo{title}{Pacemaker activity in a sensory ending with multiple
  encoding sites: the cat muscle spindle primary ending.}
\newblock \emph{\bibinfo{journal}{The Journal of Physiology}}
  \textbf{\bibinfo{volume}{498}}, \bibinfo{pages}{177--199}
  (\bibinfo{year}{1997}).

\bibitem{banks2009comparative}
\bibinfo{author}{Banks, R.}, \bibinfo{author}{Hulliger, M.},
  \bibinfo{author}{Saed, H.} \& \bibinfo{author}{Stacey, M.}
\newblock \bibinfo{title}{A comparative analysis of the encapsulated end-organs
  of mammalian skeletal muscles and of their sensory nerve endings}.
\newblock \emph{\bibinfo{journal}{Journal of Anatomy}}
  \textbf{\bibinfo{volume}{214}}, \bibinfo{pages}{859--887}
  (\bibinfo{year}{2009}).

\bibitem{besson1999neurobiology}
\bibinfo{author}{Besson, J.}
\newblock \bibinfo{title}{The neurobiology of pain}.
\newblock \emph{\bibinfo{journal}{The Lancet}} \textbf{\bibinfo{volume}{353}},
  \bibinfo{pages}{1610--1615} (\bibinfo{year}{1999}).

\bibitem{lesniak2014computation}
\bibinfo{author}{Lesniak, D.~R.} \emph{et~al.}
\newblock \bibinfo{title}{Computation identifies structural features that
  govern neuronal firing properties in slowly adapting touch receptors}.
\newblock \emph{\bibinfo{journal}{Elife}} \textbf{\bibinfo{volume}{3}},
  \bibinfo{pages}{e01488} (\bibinfo{year}{2014}).

\bibitem{walsh2015mammalian}
\bibinfo{author}{Walsh, C.~M.}, \bibinfo{author}{Bautista, D.~M.} \&
  \bibinfo{author}{Lumpkin, E.~A.}
\newblock \bibinfo{title}{Mammalian touch catches up}.
\newblock \emph{\bibinfo{journal}{Current Opinion in Neurobiology}}
  \textbf{\bibinfo{volume}{34}}, \bibinfo{pages}{133--139}
  (\bibinfo{year}{2015}).

\bibitem{lee2014sensory}
\bibinfo{author}{Lee, L.-Y.} \& \bibinfo{author}{Yu, J.}
\newblock \bibinfo{title}{Sensory nerves in lung and airways}.
\newblock \emph{\bibinfo{journal}{Comprehensive Physiology}}
  (\bibinfo{year}{2014}).

\bibitem{justus2016}
\bibinfo{author}{Kromer, J.~A.}, \bibinfo{author}{Schimansky-Geier, L.} \&
  \bibinfo{author}{Neiman, A.~B.}
\newblock \bibinfo{title}{Emergence and coherence of oscillations in star
  networks of stochastic excitable elements}.
\newblock \emph{\bibinfo{journal}{Physical Review E}}
  \textbf{\bibinfo{volume}{93}}, \bibinfo{pages}{042406}
  (\bibinfo{year}{2016}).

\bibitem{ermentrout2010foundations}
\bibinfo{author}{Ermentrout, B.} \& \bibinfo{author}{Terman, D.~H.}
\newblock \emph{\bibinfo{title}{Foundations of Mathematical Neuroscience}}
  (\bibinfo{publisher}{Springer Berlin}, \bibinfo{year}{2010}).

\bibitem{stemmler1996}
\bibinfo{author}{Stemmler, M.}
\newblock \bibinfo{title}{A single spike suffices: the simplest form of
  stochastic resonance in model neurons}.
\newblock \emph{\bibinfo{journal}{Network: Computation in Neural Systems}}
  \textbf{\bibinfo{volume}{7}}, \bibinfo{pages}{687--716}
  (\bibinfo{year}{1996}).

\bibitem{cover2012}
\bibinfo{author}{Cover, T.~M.} \& \bibinfo{author}{Thomas, J.~A.}
\newblock \emph{\bibinfo{title}{Elements of Information Theory}}
  (\bibinfo{publisher}{John Wiley \& Sons}, \bibinfo{year}{2012}).

\bibitem{ko2014}
\bibinfo{author}{Kouvaris, N.~E.}, \bibinfo{author}{Isele, T.},
  \bibinfo{author}{Mikhailov, A.~S.} \& \bibinfo{author}{Sch{\"o}ll, E.}
\newblock \bibinfo{title}{Propagation failure of excitation waves on trees and
  random networks}.
\newblock \emph{\bibinfo{journal}{EPL (Europhysics Letters)}}
  \textbf{\bibinfo{volume}{106}}, \bibinfo{pages}{68001}
  (\bibinfo{year}{2014}).

\bibitem{vankampen1985elimination}
\bibinfo{author}{Van~Kampen, N.~G.}
\newblock \bibinfo{title}{Elimination of fast variables}.
\newblock \emph{\bibinfo{journal}{Physics Reports}}
  \textbf{\bibinfo{volume}{124}}, \bibinfo{pages}{69--160}
  (\bibinfo{year}{1985}).

\bibitem{da2001explicit}
\bibinfo{author}{Da~Fonseca, C.} \& \bibinfo{author}{Petronilho, J.}
\newblock \bibinfo{title}{Explicit inverses of some tridiagonal matrices}.
\newblock \emph{\bibinfo{journal}{Linear Algebra and its Applications}}
  \textbf{\bibinfo{volume}{325}}, \bibinfo{pages}{7--21}
  (\bibinfo{year}{2001}).

\bibitem{seung1993}
\bibinfo{author}{Seung, H.~S.} \& \bibinfo{author}{Sompolinsky, H.}
\newblock \bibinfo{title}{Simple models for reading neuronal population codes}.
\newblock \emph{\bibinfo{journal}{Proceedings of the National Academy of
  Sciences}} \textbf{\bibinfo{volume}{90}}, \bibinfo{pages}{10749--10753}
  (\bibinfo{year}{1993}).

\bibitem{brunel1998}
\bibinfo{author}{Brunel, N.} \& \bibinfo{author}{Nadal, J.-P.}
\newblock \bibinfo{title}{Mutual information, fisher information, and
  population coding}.
\newblock \emph{\bibinfo{journal}{Neural Computation}}
  \textbf{\bibinfo{volume}{10}}, \bibinfo{pages}{1731--1757}
  (\bibinfo{year}{1998}).

\bibitem{pitkow2015}
\bibinfo{author}{Pitkow, X.}, \bibinfo{author}{Liu, S.},
  \bibinfo{author}{Angelaki, D.~E.}, \bibinfo{author}{DeAngelis, G.~C.} \&
  \bibinfo{author}{Pouget, A.}
\newblock \bibinfo{title}{How can single sensory neurons predict behavior?}
\newblock \emph{\bibinfo{journal}{Neuron}} \textbf{\bibinfo{volume}{87}},
  \bibinfo{pages}{411--423} (\bibinfo{year}{2015}).

\bibitem{greenwood2016}
\bibinfo{author}{Greenwood, P.~E.} \& \bibinfo{author}{Ward, L.~M.}
\newblock \emph{\bibinfo{title}{Stochastic Neuron Models}},
  vol.~\bibinfo{volume}{1} (\bibinfo{publisher}{Springer},
  \bibinfo{year}{2016}).

\bibitem{gammaitoni1998stochastic}
\bibinfo{author}{Gammaitoni, L.}, \bibinfo{author}{H{\"a}nggi, P.},
  \bibinfo{author}{Jung, P.} \& \bibinfo{author}{Marchesoni, F.}
\newblock \bibinfo{title}{Stochastic resonance}.
\newblock \emph{\bibinfo{journal}{Reviews of Modern Physics}}
  \textbf{\bibinfo{volume}{70}}, \bibinfo{pages}{223} (\bibinfo{year}{1998}).

\bibitem{PRLpikovsky97}
\bibinfo{author}{Pikovsky, A.~S.} \& \bibinfo{author}{Kurths, J.}
\newblock \bibinfo{title}{Coherence resonance in a noise-driven excitable
  system}.
\newblock \emph{\bibinfo{journal}{Physical Review Letters}}
  \textbf{\bibinfo{volume}{78}}, \bibinfo{pages}{775--778}
  (\bibinfo{year}{1997}).

\bibitem{Pikovsky2002}
\bibinfo{author}{Pikovsky, A.}, \bibinfo{author}{Zaikin, A.} \&
  \bibinfo{author}{de~La~Casa, M.~A.}
\newblock \bibinfo{title}{System size resonance in coupled noisy systems and in
  the ising model}.
\newblock \emph{\bibinfo{journal}{Physical Review Letters}}
  \textbf{\bibinfo{volume}{88}}, \bibinfo{pages}{050601}
  (\bibinfo{year}{2002}).

\bibitem{drmota2009random}
\bibinfo{author}{Drmota, M.}
\newblock \emph{\bibinfo{title}{Random Trees: An Interplay Between
  Combinatorics and Probability}} (\bibinfo{publisher}{Springer Science \&
  Business Media}, \bibinfo{year}{2009}).

\end{thebibliography}

\begin{thebibliography}{1}
\expandafter\ifx\csname url\endcsname\relax
  \def\url#1{\texttt{#1}}\fi
\expandafter\ifx\csname urlprefix\endcsname\relax\def\urlprefix{URL }\fi
\providecommand{\bibinfo}[2]{#2}
\providecommand{\eprint}[2][]{\url{#2}}

\bibitem{mc2002}
\bibinfo{author}{McIntyre, C.~C.}, \bibinfo{author}{Richardson, A.~G.} \&
  \bibinfo{author}{Grill, W.~M.}
\newblock \bibinfo{title}{Modeling the excitability of mammalian nerve fibers:
  influence of afterpotentials on the recovery cycle}.
\newblock \emph{\bibinfo{journal}{Journal of Neurophysiology}}
  \textbf{\bibinfo{volume}{87}}, \bibinfo{pages}{995--1006}
  (\bibinfo{year}{2002}).

\bibitem{hh}
\bibinfo{author}{Hodgkin, A.~L.} \& \bibinfo{author}{Huxley, A.~F.}
\newblock \bibinfo{title}{A quantitative description of membrane current and
  its application to conduction and excitation in nerve}.
\newblock \emph{\bibinfo{journal}{The Journal of Physiology}}
  \textbf{\bibinfo{volume}{117}}, \bibinfo{pages}{500} (\bibinfo{year}{1952}).

\bibitem{fh1964}
\bibinfo{author}{Frankenhaeuser, B.} \& \bibinfo{author}{Huxley, A.}
\newblock \bibinfo{title}{The action potential in the myelinated nerve fibre of
  xenopus laevis as computed on the basis of voltage clamp data}.
\newblock \emph{\bibinfo{journal}{The Journal of Physiology}}
  \textbf{\bibinfo{volume}{171}}, \bibinfo{pages}{302} (\bibinfo{year}{1964}).

\bibitem{da2001explicit}
\bibinfo{author}{Da~Fonseca, C.} \& \bibinfo{author}{Petronilho, J.}
\newblock \bibinfo{title}{Explicit inverses of some tridiagonal matrices}.
\newblock \emph{\bibinfo{journal}{Linear Algebra and its Applications}}
  \textbf{\bibinfo{volume}{325}}, \bibinfo{pages}{7--21}
  (\bibinfo{year}{2001}).

\end{thebibliography}
\end{document}